\documentstyle[12pt,cite,epsf]{article}

\catcode`\@=11
\long\def\@makefntext#1{
\protect\noindent \hbox to 3.2pt {\hskip-.9pt
$^{{\ninerm\@thefnmark}}$\hfil}#1\hfill}		

\def\@makefnmark{\hbox to 0pt{$^{\@thefnmark}$\hss}}  

\def\ps@myheadings{\let\@mkboth\@gobbletwo
\def\@oddhead{\hbox{}
\rightmark\hfil\ninerm\thepage}
\def\@oddfoot{}\def\@evenhead{\ninerm\thepage\hfil
\leftmark\hbox{}}\def\@evenfoot{}
\def\sectionmark##1{}\def\subsectionmark##1{}}

\renewcommand{\thefootnote}{\fnsymbol{footnote}}

\newcounter{sectionc}\newcounter{subsectionc}\newcounter{subsubsectionc}
\renewcommand{\section}[1] {\vspace*{0.6cm}\addtocounter{sectionc}{1}
\setcounter{subsectionc}{0}\setcounter{subsubsectionc}{0}\noindent
	{\normalsize\bf\thesectionc. #1}\par\vspace*{0.4cm}}
\renewcommand{\subsection}[1] {\vspace*{0.6cm}\addtocounter{subsectionc}{1}
	\setcounter{subsubsectionc}{0}\noindent
	{\normalsize\it\thesectionc.\thesubsectionc. #1}\par\vspace*{0.4cm}}
\renewcommand{\subsubsection}[1]
{\vspace*{0.6cm}\addtocounter{subsubsectionc}{1}
	\noindent {\normalsize\rm\thesectionc.\thesubsectionc.\thesubsubsectionc.
	#1}\par\vspace*{0.4cm}}

\newcounter{appendixc}
\newcounter{subappendixc}[appendixc]
\newcounter{subsubappendixc}[subappendixc]

\renewcommand{\appendix}[1] {\vspace*{0.6cm}
        \refstepcounter{appendixc}
        \setcounter{figure}{0}
        \setcounter{table}{0}
        \setcounter{equation}{0}
        \renewcommand{\thefigure}{\Alph{appendixc}.\arabic{figure}}
        \renewcommand{\thetable}{\Alph{appendixc}.\arabic{table}}
        \renewcommand{\theappendixc}{\Alph{appendixc}}
        \renewcommand{\theequation}{\Alph{appendixc}.\arabic{equation}}
        \noindent{\bf Appendix \theappendixc #1}\par\vspace*{0.4cm}}

\def\abstracts#1{{

\centering{\begin{minipage}{12.2truecm}\footnotesize\baselineskip=12pt\noindent
	\centerline{\footnotesize ABSTRACT}\vspace*{0.3cm}
	\parindent=0pt #1
	\end{minipage}}\par}}

\renewenvironment{thebibliography}[1]
	{\begin{list}{\arabic{enumi}.}
	{\usecounter{enumi}\setlength{\parsep}{0pt}
\setlength{\leftmargin 1.25cm}{\rightmargin 0pt}
	 \setlength{\itemsep}{0pt} \settowidth
	{\labelwidth}{#1.}\sloppy}}{\end{list}}

\newcounter{itemlistc}
\newcounter{romanlistc}
\newcounter{alphlistc}
\newcounter{arabiclistc}

\newcommand{\fcaption}[1]{
        \refstepcounter{figure}
        \setbox\@tempboxa = \hbox{\footnotesize Fig.~\thefigure. #1}
        \ifdim \wd\@tempboxa > 6in
           {\begin{center}
        \parbox{6in}{\footnotesize\baselineskip=12pt Fig.~\thefigure. #1}
            \end{center}}
        \else
             {\begin{center}
             {\footnotesize Fig.~\thefigure. #1}
              \end{center}}
        \fi}

\newcommand{\tcaption}[1]{
        \refstepcounter{table}
        \setbox\@tempboxa = \hbox{\footnotesize Table~\thetable. #1}
        \ifdim \wd\@tempboxa > 6in
           {\begin{center}
        \parbox{6in}{\footnotesize\baselineskip=12pt Table~\thetable. #1}
            \end{center}}
        \else
             {\begin{center}
             {\footnotesize Table~\thetable. #1}
              \end{center}}
        \fi}

\def\@citex[#1]#2{\if@filesw\immediate\write\@auxout
	{\string\citation{#2}}\fi
\def\@citea{}\@cite{\@for\@citeb:=#2\do
	{\@citea\def\@citea{,}\@ifundefined
	{b@\@citeb}{{\bf ?}\@warning
	{Citation `\@citeb' on page \thepage \space undefined}}
	{\csname b@\@citeb\endcsname}}}{#1}}

\newif\if@cghi
\def\cite{\@cghitrue\@ifnextchar [{\@tempswatrue
	\@citex}{\@tempswafalse\@citex[]}}
\def\citelow{\@cghifalse\@ifnextchar [{\@tempswatrue
	\@citex}{\@tempswafalse\@citex[]}}
\def\@cite#1#2{{$\null^{#1}$\if@tempswa\typeout
	{IJCGA warning: optional citation argument
	ignored: `#2'} \fi}}

 1
 1
 1

\font\ninerm=cmr9

\def\mafigura#1#2#3#4{
 \begin{figure}[hbtp]
    \begin{center}
      \epsfxsize=#1 \leavevmode \epsffile{#2}
    \end{center}
    \caption{#3}
    \label{#4}
  \end{figure} }

\def\Zslash{\hbox{$Z$\kern-0.5em\raise0.3ex\hbox{/}}}
\def\Wslash{\hbox{$W$\kern-0.5em\raise0.3ex\hbox{/}}}
\def\gt{c_L}

\newcommand{\mysection}[1]{\setcounter{equation}{0}\section{#1}}
\renewcommand{\theequation}{\thesection.\arabic{equation}}

\newcommand{\beq}{\begin{equation}}
\newcommand{\eeq}{\end{equation}}
\newcommand{\bea}{\begin{eqnarray}}
\newcommand{\eea}{\end{eqnarray}}

\newcommand{\bl}{b_L}
\newcommand{\bbl}{\overline{b}_L}

\newcommand{\msb}{$\overline{\mathrm{MS}}$ }
\newcommand{\zbb}{Z\rightarrow b\bar{b}}

\newcommand{\op}{{\cal O}}

\def\N{N_c}

\def\a{\alpha_s}
\def\mt{m_t}
\def\gf{G_F}
\def\G{{M_Z^3\sqrt 2 \gf\over 48 \pi}}
\def\4pi{(4 \pi)}
\def\dslash{\hbox{$\partial$\kern-0.5em\raise0.3ex\hbox{/}}}
\def\Dslash{\hbox{$D$\kern-0.5em\raise0.3ex\hbox{/}}}
\def\Aslash{\hbox{$A$\kern-0.5em\raise0.3ex\hbox{/}}}
\def\Gslash{\hbox{$G$\kern-0.5em\raise0.3ex\hbox{/}}}
\def\pslash{\hbox{$p\ $\kern-0.5em\raise0.3ex\hbox{/}}}
\def\m{\mu}
\begin{document}
\rightline{CERN-TH/95-92}
\vspace{2cm}

\centerline{L$\bigodot\bigodot$KING AT THE QCD CORRECTIONS}
\centerline{ FOR LARGE $M_t$:}
\centerline{AN EFFECTIVE LAGRANGIAN POINT OF VIEW\footnote{Talk given at the
Ringberg Workshop ``Perspectives for
electroweak interactions in $e^+e^-$ collisions", Ringberg Castle, Germany,
February 5--8, 1995. To appear in the proceedings.}}
\vskip 1cm
\centerline{S. Peris\footnote{On leave from
Grup de F\'{\i}sica Teorica and IFAE,
Universitat Autonoma de Barcelona, Barcelona, Spain.
peris@surya11.cern.ch}}
\centerline{TH Division, CERN, 1211 Geneva 23, Switzerland}

\vspace{2cm}

\abstracts{We discuss the QCD
corrections to the large-$m_t$ electroweak
contributions to $\Delta r$ and to the process $Z\to b \bar b$ as two of the
most representative examples. This needs the construction of an effetive
field theory below the top quark. We discuss the issue of what $\mu$ scale is
the appropriate one at every stage  and argue that, while matching
corrections do verify the simple prescription of taking $\mu \simeq m_t$ in
$\alpha_s(\mu)$, logarithmic (i.e. $\sim \log m_t$) corrections do not, and
require the use of the running $\alpha_s(\mu)$ in the corresponding
renormalization group equations. In particular we obtain the $\alpha_s$
correction to the non-universal $\log m_t$ contribution to the $Zb\bar b$
vertex.}

\vspace{0.5cm}
\vfill\noindent
CERN-TH/95-92

\noindent

\newpage

\textwidth 6.0in
\textheight 8.6in
\pagestyle{empty}
\topmargin -0.25truein
\oddsidemargin 0.30truein
\evensidemargin 0.30truein
\parindent=1.5pc
\baselineskip=15pt

\centerline{\normalsize\bf L$\bigodot\bigodot$KING AT THE QCD CORRECTIONS}
\baselineskip=16pt
\centerline{\normalsize\bf FOR LARGE $M_t$ :}
\baselineskip=22pt
\centerline{\normalsize\bf AN EFFECTIVE FIELD THEORY POINT OF VIEW}

\centerline{\footnotesize S. PERIS\footnote{On leave from Grup de Fisica
Teorica and IFAE, Universitat Autonoma de Barcelona, Barcelona, Spain.}}

\baselineskip=13pt
\centerline{\footnotesize\it Theory Division, CERN}
\baselineskip=12pt
\centerline{\footnotesize\it CH-1211 Geneva 23, Switzerland}
\centerline{\footnotesize E-mail: peris@surya11.cern.ch}

\vspace*{0.9cm}
\abstracts{We discuss the QCD corrections to the large-$m_t$ electroweak
contributions to $\Delta r$ and to the process $Z\to b \bar b$ as two of the
most representative examples. This needs the construction of an effective
field theory below the top quark. We discuss the issue of what $\mu$ scale is
the appropriate one at every stage  and argue that, while matching
corrections do verify the simple prescription of taking $\mu \simeq m_t$ in
$\alpha_s(\mu)$, logarithmic (i.e. $\sim \log m_t$) corrections do not, and
require the use of the running $\alpha_s(\mu)$ in the corresponding
renormalization group equations. In particular we obtain the $\alpha_s$
correction to the non-universal $\log m_t$ contribution to the $Zb\bar b$
vertex.}

\normalsize\baselineskip=15pt
\setcounter{footnote}{0}
\renewcommand{\thefootnote}{\alph{footnote}}
\mysection{Introduction}

\setcounter{section}{1}

Electroweak (EW)
radiative corrections are presently achieving an
extremely high degree of
sophistication and complexity. After the high-precision experiments
recently performed
at LEP and the SLC \cite{Miquel} there is a clear need for increasingly
higher-order calculations,
even if only
for assessing the size of the theoretical error when comparing to the
experiment. Currently two-loop EW corrections (pure or mixed with
QCD) are being analyzed rather systematically \cite{Kniehl-report}
and, sometimes, even up to three loops are being
accomplished \cite{Tarasov-rho}.
Needless to say these calculations are
extremely complicated
and usually heavily rely on the use of the computer.
In this paper we would like to point out that in some
situations thinking in terms of
effective field theories (EFTs) \cite{efts,efts2} can help in this
development.

Built as a systematic approximation scheme for problems with widely
separated scales \cite{Hall}, EFTs organize the calculation
in a transparent
way dealing with
one scale at a time and clearly separating the physics of the
ultraviolet from the physics of the infrared. They are based on the
observation that, instead of obtaining the full answer and then taking
the appropriate interesting limits, a more efficient strategy  consists
in taking the limit first, whereby considerably reducing the amount of
complexity one has to deal with, right from the start.
For this kind of problems EFTs are never more
complicated than the actual loop-wise perturbative
calculation and in some
specific cases they may even be more advantageous, being even
able to render an extremely
complicated calculation something very simple.

By EFT we specifically mean the systematic construction of the effective
Lagrangian that results when a heavy particle is integrated out. The
procedure goes as follows \cite{efts,efts2}. Let us
imagine we are interested in studying the
physics at an energy scale $E_0$. Starting at a scale $\mu>>E_0$ one uses
the powerful machinery of the renormalization group equations (RGEs) to
scale the initial Lagrangian from the scale $\mu$ down to the energy $E_0$
one is interested in. If in doing so one encounters a certain particle with
mass $m$, one must integrate this particle out and find the corresponding
matching conditions so that the physics below and above the scale $\mu=m$
(that is to say the physics described by the Lagrangian with and without the
heavy particle in question) is the same. This is technically achieved by
equating
the one-particle irreducible Green functions (with respect to the other
light fields) in both theories to a certain order in inverse powers of the
heavy mass $m$.\footnote{One could also match S-matrix elements.}\ This
usually requires the introduction of local counterterms \cite{Witten} in the
effective Lagrangian for $\mu<m$. Once this is done, one keeps using the
RGEs
 until the energy $E_0$ is reached. If another particle's threshold is
crossed, the above matching has to be performed again. All this procedure is
most efficiently carried out by using the \msb renormalization scheme
, where the RGEs are mass-independent and can be gotten directly from
the $1/\epsilon$ poles of dimensional regularization. Schematically, the
standard strategy in the case of the top quark is the following:

\begin{enumerate}
\item Matching the effective theory to the full theory at $m_t$.
\item Running the effective Lagrangian from $m_t$ down to $M_Z$.
\item Calculating matrix elements with the effective Lagrangian at the
scale $M_Z$.
\end{enumerate}

\mysection{$\Delta r$}

\setcounter{section}{2}

According to refs. \cite{Djouadi,Sergio,Jegerlehner,yellow},
the top contribution to $\Delta r$ can be expressed as
\footnote{There is a typographical error in the $\log M^2_W/m_t^2$
term of $\Delta r(top)$ in ref. \cite{yellow}, which appears with an
overall minus
sign with respect to our expression. We thank F. Jegerlehner for confirming
this.}

$$
\Delta r(top)\approx -{c^2\over s^2}\ {3 \mt^2 \gf \sqrt 2\over \4pi^2}
\left(1-{\a(\mu)\over \pi}{6+2 \pi^2\over 9}\right)+
$$

\beq
\label{A0}
\qquad +{g^2\over \4pi^2}\ {1\over 2} \left({c^2\over s^2}-{1\over 3}\right)
\log \left({M_W^2\over \mt^2}\right)
\left(1+ {\a(\mu)\over \pi}\right)\qquad ,
\eeq
where we have only kept the leading and next-to-leading $\mt$ dependence.

Along with these results, there has appeared the discussion of the $\mu$
scale at which one
is supposed to evaluate $\a(\mu)$ in these expressions, \footnote{In principle
three scales appear in these loops: $\mt,M_Z,m_b$.}\ and
the parameters in terms of which
one ought to express the result, i.e. whether \msb, or
on-shell, etc. For this, a prescription has been designed that
says\cite{Sergio} that corrections coming from the $(t,b)$ doublet
should be
computed with $\a(\mt)$. This prescription would then say that in all of the
above
expressions $\a(\mu)$ should be taken as $\a(\mt)$. We would like to explain
what an effective field theory (EFT) point of view shows about this
issue.
We shall see that while
this prescription works for the power-like terms
(those that go like $\mt^2$),
the renormalization group (RG) supplies us with a different result for the
logarithmic terms (those that go like $\log \mt$). \footnote{From an effective
field theory point of view these
two types of contributions are totally different. While the former (i.e.
power-like) comes
from ``matching", the latter (i.e. logarithmic) comes from ``running".
See below.}

Here we shall be concerned with the QCD corrections to the
large-$m_t$ one-loop electroweak corrections. Therefore, for all practical
purposes, one may think as if the top quark were the heaviest particle in
the SM, much heavier than the Higgs boson, which is taken to be
nearly degenerate with the W and
Z. This automatically kills the $\log M_H/M_W$ contributions and leaves the
$\mt^2$ and the $\log \mt/M_W$ ones, which are those we are interested in.

The general philosophy will parallel that so successfully used in the
context of grand unified theories\cite{Hall}. There is of course a very
important  difference, namely that, upon
integration of the top quark, the resulting effective theory will no longer
exhibit an explicit linear SU$_2\times$U$_1$ invariance\cite{efts2}.
This would make a full
account of the corresponding RGEs very cumbersome. Luckily we may
keep only those contributions that are strictly relevant.

\mafigura{10cm}{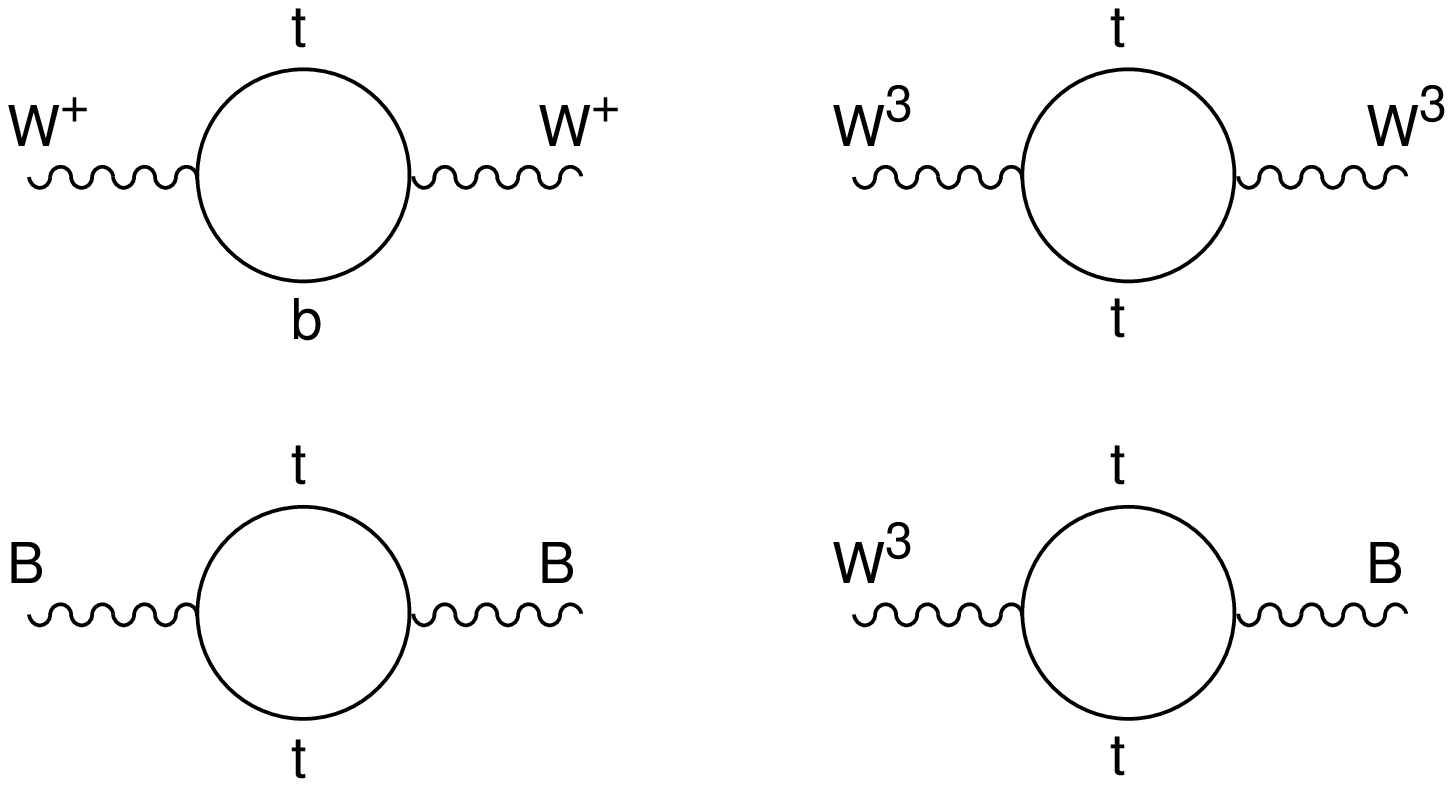}{Diagrams contributing to the matching
conditions, eqs. (\ref{five0}-\ref{six0}).}{zbb0}

Let us start with the full SM at $\mu>\mt$. At $\mu=\mt$, one
integrates the top out, obtaining\footnote{Because of lack of space the
reader interested in the details regarding this section is referred to ref.
\cite{Peris}.}

$${\cal L}= W_\mu^+ \partial^2 {W^-}^\mu+
{g_+^2(\m)\over 4} \left( v^2 +
\delta v_+^2(\mu)\right) W^+_\mu {W^-}^\mu +
{1\over 2} W^3_\mu \partial^2 {W^3}^\mu + {1\over 2} B_\mu \partial^2 B^\mu
+
$$

$$
\qquad +{1\over 2} \left(g_3(\mu) W_3^\mu - g'(\mu)  B^\mu \right)
\left[{1\over 4}
\left(v^2+\delta v^2_3(\mu)\right)-\delta Z_{3Y}(\mu) \partial^2 \right]
\left(g_3(\mu) {W_3}_\mu - g'(\mu) B_\mu \right)
$$

\beq
\label{two0}
\qquad +\ \bar \psi \ i
\Dslash(\ g_+(\mu)W^+,\ g_3(\mu)W_3,\ g'(\mu)B)\ \psi \ ,
\eeq
from the diagrams of fig. \ref{zbb0} after a trivial field redefinition.
Here $\psi$ stands
for all the fermions but the
top. Notice that we have dealt with $W_3-B$ mixing by including a
$\partial^2$ operator in the form of a ``mass term" in eq.
(\ref{two0}).
This will make the subsequent diagonalization very simple since the neutral
mass eigenstate is still of the form $gW_3-g'B$, like at tree level.
Certainly, there will also be a tower of
higher dimensional operators suppressed by the corresponding inverse
powers of the
top quark mass, but we shall neglect them. Possible four-fermion
operators are irrelevant to the discussion that follows and are also
disregarded.
We also postpone the study of the $Zb\bar b$ vertex to the next sections.

To the order we are working, i.e. one loop:
$$
g_+^2(\mu)\approx g^2 \left(1-g^2 \delta Z_+(\mu)\right)
$$

$$
g_3^2(\mu)\approx g^2\left(1-g^2 \delta Z_3(\mu)-g^2\delta
Z_{3Y}(\mu)\right)
$$

\beq
\label{three0}
g'^2(\mu)\approx g'^2\left( 1-g'^2 \delta Z_Y(\mu) -g'^2 \delta
Z_{3Y}(\mu)\right)\qquad .
\eeq

Notice that below the top quark mass the initially unique coupling constant
$g$ has split into $g_+$ and $g_3$ \cite{Peccei}.
Similarly $v_+^2(\mu)\equiv v^2 +
\delta v^2_+(\mu)$ and $v^2_3(\mu)\equiv v^2+\delta v_3^2(\mu)$ are also
different. The matching
conditions are very easily obtained sine they are nothing else than the
diagrams of fig. 1 evaluated at $\m=\mt$. This means that
\beq
\label{five0}
\delta v_+^2(\mt)={N_c\over \4pi^2}\ \mt^2\qquad ,\qquad
\delta {v_3}^2(\mt)=0\quad .
\eeq

Analogously,
\beq
\label{six0}
\delta Z_{3Y}(\mt)=0\ ,\quad g_+(\mt)=g_3(\mt)=g\ \ {\rm and}\ \
g'(\mt)=g'\quad .
\eeq
Equation (\ref{six0}) says that the coupling
constants are continuous across the
threshold. This is true as long as one keeps only the leading logarithms. In
general there are non-logarithmic pieces that modify (\ref{six0}) such as, for
instance, the non-log term in the first of eqs. (\ref{five0}).
The point is that
this term in (\ref{five0}) is multiplied by $\mt^2$
(i.e. a large non-decoupling effect)
and therefore contributes  (in fact dominates) for large $\mt$, whereas
the same does not happen in (\ref{six0}). Therefore, non-log
corrections to (\ref{six0}) do not affect
the large-$\mt$ discussion that follows.

In order to obtain the effective Lagrangian at the relevant lower scales
$\m\simeq M\equiv M_W, M_Z$\footnote{Hence we neglect possible
terms $\sim \log
M_W/M_Z$.} one has to scale this Lagrangian down using the RGE for
each ``coupling" $g_+(\m),g_3(\m),g'(\m),\delta v_+^2(\m), \delta v_3^2(\m)$
and $\delta Z_{3Y}(\m)$. The running of
$\delta v_{+,3}^2(\m)$ is zero since
it must be proportional to a light fermion mass, which we neglect.
\footnote{We
also neglect the contribution of the gauge bosons and the Higgs since they
do not have QCD corrections. This simplifies the analysis enormously.}
Therefore,
\beq
\label{seven0}
\delta v_{+,3}^2(\mt)=\delta v_{+,3}^2(M)\qquad .
\eeq

After including $\a$ corrections, one immediately
obtains ($t\equiv \log \m^2$)\cite{Peris,Jones},

$$
{dg^2_+\over dt}={g^4_+\over \4pi^2} \left[2\ \left(1 +{\a
(t)\over \pi}\right) + 1\right] +...
$$

$$
{dg^2_3\over dt}={g^4_3\over \4pi^2} \left[{7\over 3}\ \left(1 +{\a
(t)\over \pi}\right) + 1\right] +...
$$

$$
{dg'^2\over dt}={g'^4\over \4pi^2} \left[{23\over 9}\ \left(1 +{\a
(t)\over \pi}\right) + 3\right] +...
$$

\beq
\label{ten0}
{d\over dt} \delta Z_{3Y}={1\over 6\4pi^2} \left(1 +{\a
(t)\over \pi}\right) +...
\eeq
from the diagrams of fig. \ref{zbb0}, but with gluon corrections.
Ellipses in eq.
(\ref{ten0}) stand for the contribution of the gauge bosons and the
Higgs.\footnote{We note again that this contribution will not have QCD
corrections to the order we
are working.}\ ${\cal O}(\a)$ corrections do not affect eqs.
(\ref{six0}),(\ref{seven0}). Equations (\ref{ten0}) are to be
supplemented with the running of $\a(t)$,
\beq
\label{eleven0}
{d\a\over dt}=-{\beta_0\over \4pi}\ \a^2\quad ,\quad \beta_0=11-{2\over 3} n_f\
,\
n_f=5\ {\rm flavors}\quad .
\eeq

The boundary conditions (\ref{five0}) read\cite{Djouadi}

$$
{v_+^2(M)-v^2_3(M)\over v^2_+(M)}=
{v_+^2(\mt)-v^2_3(\mt)\over v^2_+(\mt)}=\qquad \qquad \qquad
$$

\beq
\label{nine0}
\qquad ={3\over \4pi^2}{\mt^2(\mt)\over v^2_+(\mt)}\ \left[1- {2\over
9}{\a(\mt)\over \pi}(\pi^2-9) + {\cal O}(\a^2)\right]\quad ,
\eeq
where, as nicely explained in refs.
\cite{Grinstein-Wang,Georgi-Cohen-Grinstein} , the scale in
$\a(\m)$ and
$\mt(\m)$ clearly has to be $\sim \mt$ (and not $M_Z$ or $m_b$) because it is
nothing but a matching condition at $\m=\mt$. This is the $\rho$ parameter.
We shall see below that $v^2_+(\mt)=(\sqrt 2 \gf)^{-1}$, where $\gf$ is the
$\m$-decay constant. Recently Sirlin\cite{Sirlinone} has noted certains
virtues in an expression like eq. (\ref{nine0}).
Within the EFT approach it comes out very naturally.

Given that $g_+^2,g_3^2$ and $g'^2$ are all rather smaller than $g_s^2\equiv
4\pi \a$, a reasonable
approximation is to take into account the running of $\a$ in eqs.
(\ref{ten0}) while
keeping the $g_+, g_3$ and $g'$ frozen at a given value. This is tantamount
to resumming
the leading log's accompanying powers of $\a$ but not those accompanied by
powers of $g_+, g_3$ and $g'$.

With all this, one can now go about computing a typical physical quantity
like for instance $\Delta r_W$, which is the same as the more
familiar parameter $\Delta r$ defined by Marciano and Sirlin\cite{Marciano}
but
without the running of $e(\mu)$. In the EFT language this is obtained
in the following way. According to the Lagrangian (\ref{two0})  the physical
$W$ and
$Z$ masses are given by the equations
$$
M_W^2={g^2_+(M)\over 4}\ v^2_+(M)
$$

\beq
\label{twelve0}
M_Z^2={g_3^2(M)\over 4 c^2(M)}\ \left(v^2_3(M)+ 4 M_Z^2 \delta
Z_{3Y}(M)\right)\quad ,
\eeq
where $c^2(M)\equiv \cos^2\theta_W(M)$ and $\tan\theta_W(M)\equiv
g'(M)/g_3(M)$.

Following the EFT technique, at the scale of the W mass one should
integrate out the W boson. This gives rise to the appearance of 4-fermion
operators that mediate $\m$ decay, with strength $\gf(M)/\sqrt 2$.
The matching condition therefore becomes
\beq
\label{thirteen0}
{\gf(M)\over \sqrt 2}= {g^2_+(M)\over 8 M_W^2}={1\over 2
v^2_+(M)}\ ,
\eeq
but since $\gf(\m)$ does not run\cite{Fermi} one can see that actually
$v^2_+(M)=\sqrt2 \gf$, where $\gf$ is the Fermi constant as measured in
$\mu$ decay. Therefore
\beq
\label{fourteen0}
{\gf \over \sqrt2}={g^2_+(M)\over 8 M_W^2}={e^2(M)\over 8
M^2_W}\left[{g^2_+(M)\over g^2_3(M)}\ {1\over s^2(M)}\right]\ ,
\eeq
where $e^2(\m)$ is the running electromagnetic coupling constant. The
quantity $\Delta r_W$ is defined as

\beq
\label{deltar0}
{\gf\over \sqrt 2}={e^2(M)\over 8 M_W^2 s^2} \ (1+\Delta r_W)\quad .
\eeq
Consequently,
\beq
\label{fifteen0}
1+\Delta r_W = {s^2\over s^2(M)}\ {g^2_+(M)\over g_3^2(M)}\quad ,
\eeq
where $s^2\equiv 1- M^2_W/M^2_Z$ is Sirlin's combination\cite{Sirlin}.

Since we are only interested in resumming $\a$ corrections we can
approximate $\Delta r_W$ in eq. (\ref{fifteen0}) by
\beq
\label{seventeen0}
\Delta r_W\approx {c^2-s^2 \over s^2}\ {g^2_3(M)-g^2_+(M)\over g^2} -
{c^2\over s^2}{v^2_+(\mt)-v^2_3(\mt)\over v^2}+{4M^2_Z\over v^2}{c^2\over
s^2} \delta Z_{3Y}(M)\ .
\eeq

Integration of eqs. (\ref{ten0}) and (\ref{eleven0}), with the
boundary conditions (\ref{six0}), yields

$${g^2_+(M)\over g^2_3(M)}\approx 1 + {g^2\over \4pi^2}\left[ -{1\over 3}
\log \left({M^2\over \mt^2}\right)+
\log\left({\a(M)\over \a(\mt)}\right)^{-{4\over
\beta_0}}\right]
$$

\beq
\label{eighteen0}
\delta Z_{3Y}(M)\approx {1\over 6\4pi^2}\left[ \log\left({M^2\over
\mt^2}\right) + \log\left({\a(M)\over \a(\mt)}\right)^{-{4\over
\beta_0}}\right]
\eeq

so that $\Delta r_W$ is, finally,
$$
\Delta r_W \approx -{c^2\over s^2} {3\over \4pi^2} \mt^2(\mt) \gf
\sqrt 2 \left[ 1- {2\over 9}{\a(\mt)\over \pi}(\pi^2-9)\right]+
$$

\beq
\label{nineteen0}
\qquad +{g^2\over \4pi^2}\ {1\over 2}\left({^2\over s^2}-{1\over 3}\right)
\left[\log\left({M^2\over
\mt^2}\right)+\log\left({\a(M)\over \a(\mt)}\right)^{-{4\over
\beta_0}}\right]
\eeq
with $\beta_0=23/3$. In the second term of eq. (\ref{nineteen0}) one has
actually
resummed all orders in $\a^n \log^n$. It is here that the powerfulness of
the RG and EFT has proved to be very useful. Therefore we learn that
while the term proportional to
$\mt^2(\mt)$ comes from matching, and has therefore a well-defined scale
$\m\simeq \mt$; the term proportional to $g^2$ comes from running,
which in turn means that it has to depend on the two
scales between which
it is running, $\m\simeq M$ and $\m\simeq \mt$. From the point of view of an
EFT aficionado, eq. (\ref{nineteen0}) is somewhat unconventional in that
it considers
matching conditions (the $\a(\mt)$ term) together with running (the
$\a(M)/\a(\mt)$ term) both at one loop. From the QCD point of view the
former is a next-to-leading-log term whereas the latter is a leading-log
one. The reason for
taking both into account is of course that the $\a(\mt)$ term is multiplied
by the  $\mt^2\gf$ combination, which is large.

If one takes the $\a(M)/\a(\mt)$ logarithmic term, expands it in powers
of $\a$ and uses
\beq
\label{pole}
m_t\approx m_t(m_t)\left( 1 + {4\a(m_t)\over 3\pi}\right)
\eeq
to rewrite eq. (\ref{nineteen0})
in terms of the pole
mass, one of course reobtains eq. (\ref{A0})  to the given order.

%

\mysection{$Z\to b \bar b$}

\setcounter{section}{3}

The decay width $\zbb$ can be written as \cite{zbb-general,RQCD}

\begin{equation}
\label{one}
\Gamma(Z\to b \bar b)= \N \ \G \ \ \rho \ \ R_{QCD}\  R_{QED} \
\ [A^2 + V^2]\ \ ,
\end{equation}
with
\begin{equation}
\label{two}
A=1 + {1\over 2} \Delta \rho^{vertex}\qquad ; \qquad
V=1 + {1\over 2} \Delta \rho^{vertex}- {4\over 3} \kappa s_0^2\ \ ,
\end{equation}
\begin{equation}
\label{three}
\kappa \approx  1 - {c^2\over c^2-s^2}\  \Delta\rho \ + {g^2\over \4pi^2}\
\ {1\over 6\ (c^2-s^2)}\log{M_W^2\over \mt^2}\
\left(1+{\a(\m)\over \pi}\right)\ \ ,
\end{equation}
\begin{equation}
\label{four}
\rho=1 + \Delta \rho \qquad , \qquad
\Delta \rho\approx {3\over \4pi^2} \mt^2(m_t) \gf \sqrt 2
\left(1- {\a(\mu)\over \pi}\ \ {2\over 9}\ (\pi^2-9)\right) \ \ ,
\end{equation}
and
\begin{equation}
\label{five}
s_0^2={1\over 2}\left(1- \sqrt{1- {4\pi \alpha(M_Z)\over \sqrt 2 \gf
M_Z^2}}\right)\ \ ,
\end{equation}
where
\begin{equation}
\label{six}
R_{QCD}\approx 1 + {\a(\m)\over \pi}\quad , \quad R_{QED}\approx 1 +
{\alpha(\m)\over 12 \pi} \ \ ,
\end{equation}
$$
\Delta \rho^{vertex}\approx  -{4 \mt^2(\mt) \gf\sqrt 2\over
\4pi^2}\left(1 -  {\a(\m)\over \pi} \ {\pi^2 - 8\over 3}\right)+
\qquad \qquad
$$
\begin{equation}
\label{seven}
\qquad \qquad +{g^2\over \4pi^2}\ \ \log
\left({M^2_W\over \mt^2}\right)\  \left({8\over 3}+
{1\over 6 c^2}\right)\left(
1+ \ {\cal C} \ {\a(\mu)\over \pi}\right) \ \ .
\end{equation}
We employed the running \msb
$\mt(\mu=\mt)$.

Here we shall describe an effective field theory calculation of the
physical process $Z\rightarrow b \bar b$. As a result we shall obtain the
value of the coefficient ${\cal C}$ in eq. (\ref{seven}). This coefficient
has also been recently obtained in ref.~\cite{Kwiatkowski-Steinhauser}
and our result agrees with theirs. Moreover, our construction of the EFT
will also yield the value for the natural scale $\mu$ that appears in the
different terms of eqs. (\ref{three}),(\ref{seven}). Again because of
limitations of space we refer the reader to ref. \cite{Peris-Santamaria} for
any detail regarding this section.

Integrating the top quark out affects the coupling
to the $W$ and $Z$ gauge bosons of every lighter fermion
through vacuum polarization as we saw in the previous section.
Moreover it also
affects specifically the coupling of the
bottom quark to the $Z$ boson.
\mafigura{7cm}{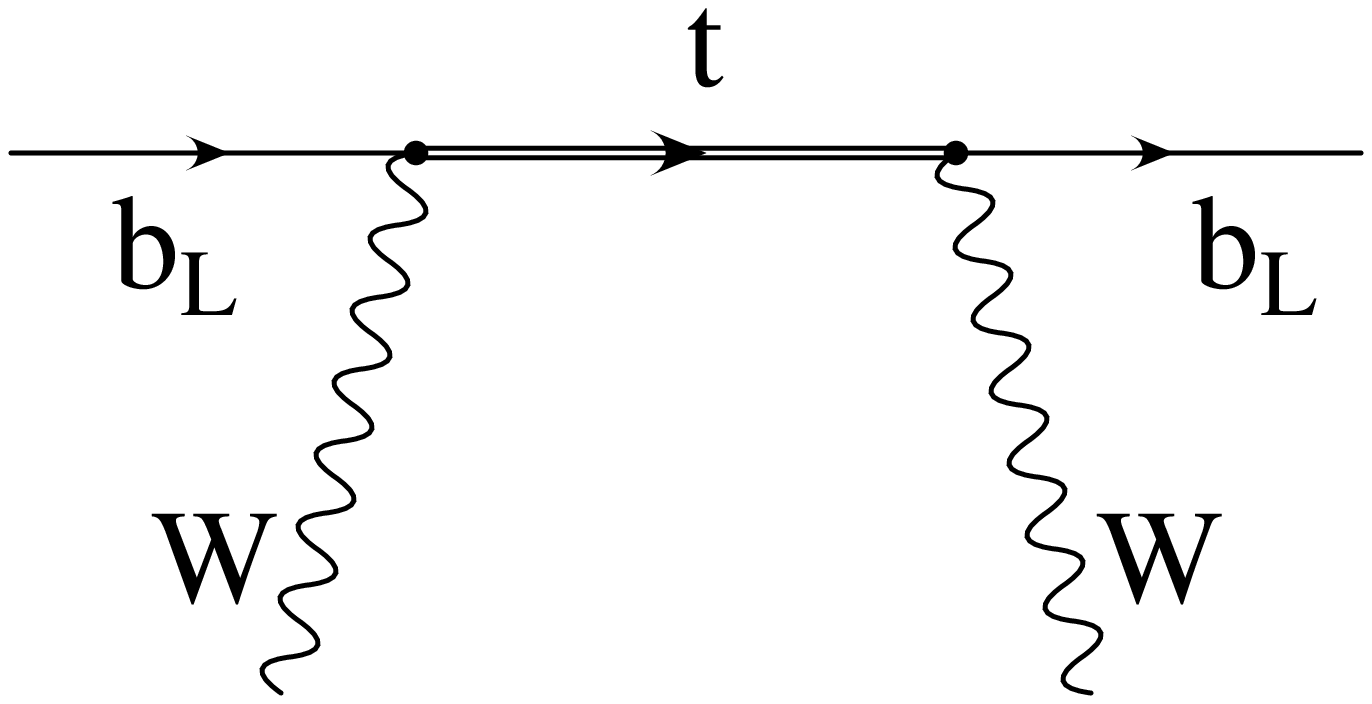}{Diagram contributing to the matching in the
unitary gauge. It is suppressed by $1/\mt^2$.}{zbb1}
The integration of the top quark is done in
several steps. Firstly, at tree
level, there is the contribution given by the diagram of fig.~\ref{zbb1}.
This contribution gives rise to an effective operator
that is suppressed by two
inverse powers of the top mass. Since ultimately
this fact is due to dimensional analysis, it cannot change once QCD is
switched on and one-loop $\a$ corrections to the diagram of fig.~\ref{zbb1}
are also considered in the matching conditions.
We shall consistently neglect this type of
contributions since they can never give rise to the terms we are
interested
in, i.e. eq. (\ref{seven}). This is the only contribution in the
unitary gauge, which is the one we shall
employ.\footnote{In the previous section we did not need to fix the gauge
since the fermion vacuum polarization is gauge invariant.} In any other
gauge other effective operators arise
because the would-be Nambu--Goldstone
bosons couple proportionally to the top
mass and may compensate the $m_t^2$
factor in the denominator.

The effective Lagrangian below the top quark mass reads:

$$
{\cal L}={\cal L}_4 + {\cal L}_6 \ \ ,
$$
$$
{\cal L}_4=\bar b\ i\Dslash\ b - \frac{1}{2}\gt^b(\m)\
\bar b\ \Zslash\ P_L\ b +
{1\over 3} c_V^b(\m)\ \bar b\ \Zslash\ b +
$$
$$
+\bar e\ i \Dslash\ e -\frac{1}{2} \gt(\m)\ \bar e\ \Zslash\ P_L\ e +
c_V(\m) \bar e\ \Zslash\ e + \frac{c_+(\m)}{\sqrt 2}
\left(\bar e\ \Wslash\ P_L\ \nu + {\rm h.c.}\right)\qquad ;
$$

\beq
\label{one2}
{\cal L}_6=\frac{1}{\Lambda_F^2}\ \sum_i\ c_i(\m) {\cal O}_i \qquad ,
\eeq
where $P_L$ is the lefthanded projector and
$\Dslash$ stands for the QED and QCD covariant derivatives. The
$c(\m)$'s of the electron are actually
common to all the fermions but the
bottom quark. For instance, the $Z\nu \bar \nu$ would be $+c_L(\m)/2$
since the neutrino has no vector coupling $c_V(\m)$.
Notice
that we have decomposed the $Zf\bar f$ vertex in terms of a lefthanded and
vector couplings instead of the more conventional left and righthanded, or
vector and axial counterparts. In eq. (\ref{one2}) $\Lambda_F=4 \pi v$,
$v=(\sqrt 2 G_F)^{-1/2}= 246$~GeV and  the ${\cal O}_i$'s are a set of
dimension-six operators involving the (lefthanded)
bottom quark and three (covariant) derivatives; or
the bottom quark, the $Z$ and two
derivatives. They arise from the longitudinal part of the $W$
propagators. This is why the scale $\Lambda_F$ appears: it is the
combination of the ordinary $1/m_t^2$ suppression of any six-dimensional
operator in an effective field theory and the fact that the would-be
Nambu--Goldstone bosons couple proportionally to the top mass.

For convenience we have changed here the notation for the effective couplings
with respect to the previous section. The connection is given by
\beq
\label{tururu}
c_L(\m)=\frac{g_3(\m)}{c(\m)}\quad ; \quad
\ c_V(\m)=\frac{g_3(\m) s^2(\m)}{c(\m)}\quad
; \quad c_+(\m)=g_+(\m)\qquad,
\eeq
where $s^2(\m)=\sin^2\theta_W(\m)$ and $\tan\theta_W(\m)=
g'(\m)/g_3(\m)$.

\mafigura{9cm}{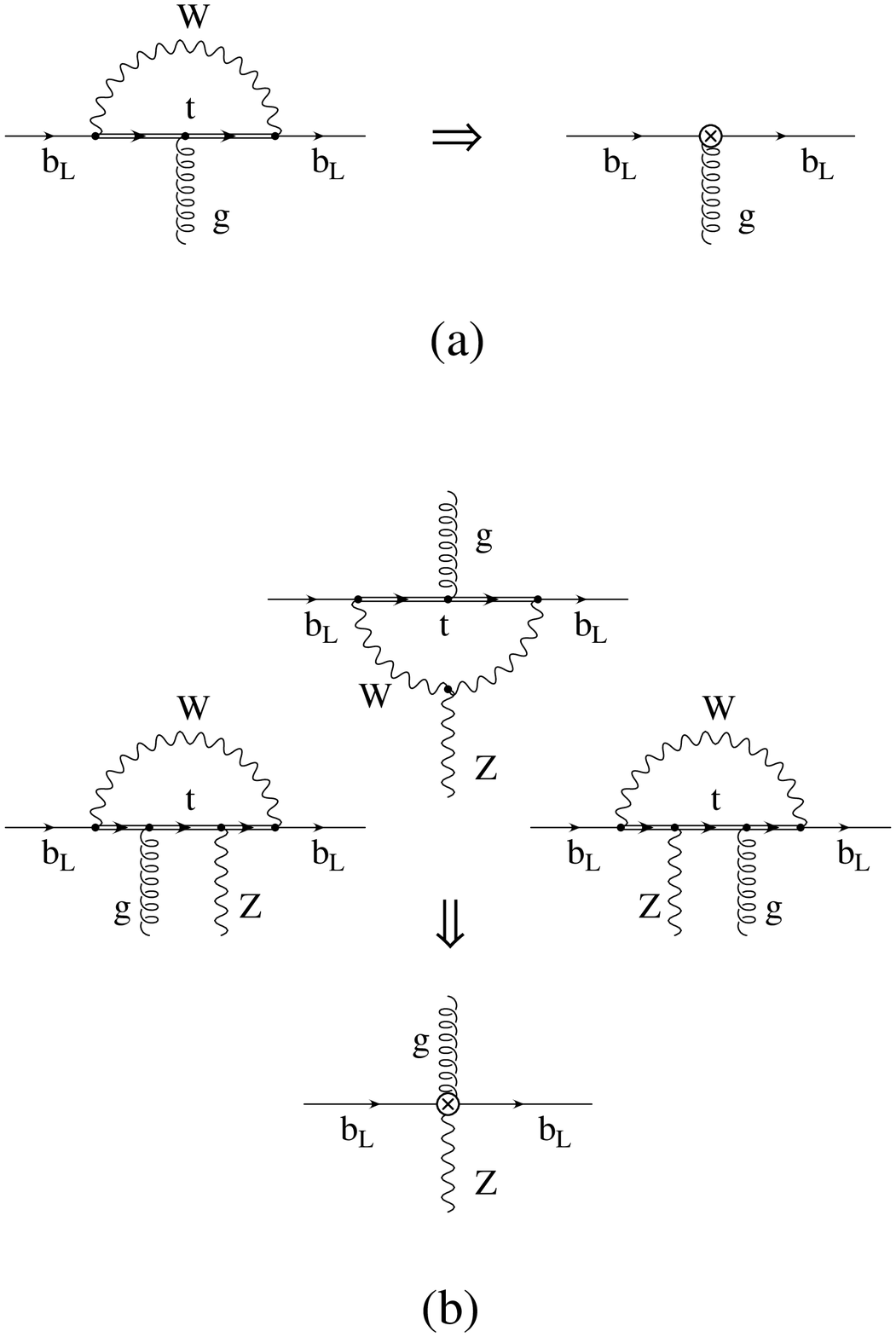}{One-loop matching due to QCD.}{zbb6}

We can select  the non-universal part of the $Zb\bar b$ vertex by comparing
the $\gt^b(\m)$ coupling on shell with the analogous coupling for the
electron $\gt(\m)$ at the scale $\m\sim M_Z\sim M_W\equiv M$. One
defines\footnote{This ratio is called $1+\epsilon_b$ in
ref.~\cite{Altarelli-Barbieri-Caravaglios-NPB405}}
\beq
\label{alta}
1+\frac{1}{2} \Delta \rho^{vertex} = \frac{\gt^b(M)}{\gt(M)}\ \ .
\eeq

In order to make contact with the physics at the scale $\m=M$, one has to
scale the Lagrangian (\ref{one2}) down to this particular $\m$.
In this process of scaling,
$\gt^b(\m)$ and $\gt(\m)$ run differently. The calculation can be
done by setting the external particles on shell.

Clearly the effect of integrating the top quark out affects only the
lefthanded projection of the bottom-quark field, i.e. $c_L^b(\m)$, but
leaves untouched the coefficient $c_V^b(\m)$. As a matter of fact
$c_V^b(\m)=c_V(\m)$.

The diagrams of
fig.~\ref{zbb6} give rise to the dimension-six
operators that appear in eq. (\ref{one2}).
In principle one should now calculate how all these operators mix back
into the $Zb\bar b$ operators of eq. (\ref{one2}) and make the
coefficients $c^b_{L,V}(\m)$ evolve with $\m$ as one runs from $m_t$
down to $M_Z$. However, a clever
use of the equations of motion\cite{Arzt} helps us get rid of almost all
the operator structures that are generated in the matching and leaves us
with only one operator that is interesting. This one is
\footnote{One could still use the equations of motion for the gluon
field but we found more convenient not to do so.}

\beq
\label{arca1}
\op_1 = \bbl \gamma^\nu \frac{\lambda^A}{2}
\bl \ g_s\ D^\mu G^A_{\mu \nu}\qquad .
\eeq
Of course this operator only affects the running of $c^b_L(\m)$ and not
of $c_V^b(\m)$.

An explicit straightforward evaluation of
the diagrams of fig.~\ref{zbb6}a yields
for the coefficient $c_1(\m)$ accompanying the operator $\op_1$ the
value
\beq
\label{arca2}
c_1(m_t)=-\frac{7}{18} \ \ .
\eeq

In order to make contact with the physics at the scale $\m \simeq M$
one has to find how $c_L^b(\m)$ scales with $\m$. We use the Feynman gauge
propagator for the gluon.
One obtains\cite{Peris-Santamaria}
\beq
\label{arca3}
\frac{dc_L^b(t)}{dt}=\left(\mathrm{something}\right)
+\ \frac{g}{c}\ \frac{g^2}{\4pi^2}\ \gamma_1\ c_1(t) \frac{\a(t)}{\pi}
\ \ \ ,
\eeq
where the second term comes from fig. \ref{zbb7}b and ``something" stands
for a certain contribution common to the running of $c_L(\m)$ that will
cancel in the final ratio (\ref{arca5}) (see below). We obtain
the following value for the coefficient $\gamma_1$:

\beq
\label{gamma1}
\gamma_1=-\frac{1}{9 c^2}\left(1-\frac{2}{3} s^2\right)\ \ .
\eeq
Since $\op_1$ only involves the lefthanded bottom quark it is clear why the
coefficient $\gamma_1$ turns out to be proportional to the lefthanded bottom
coupling to the $Z$, i.e. the combination $1-\frac{2}{3}s^2$.

\mafigura{8cm}{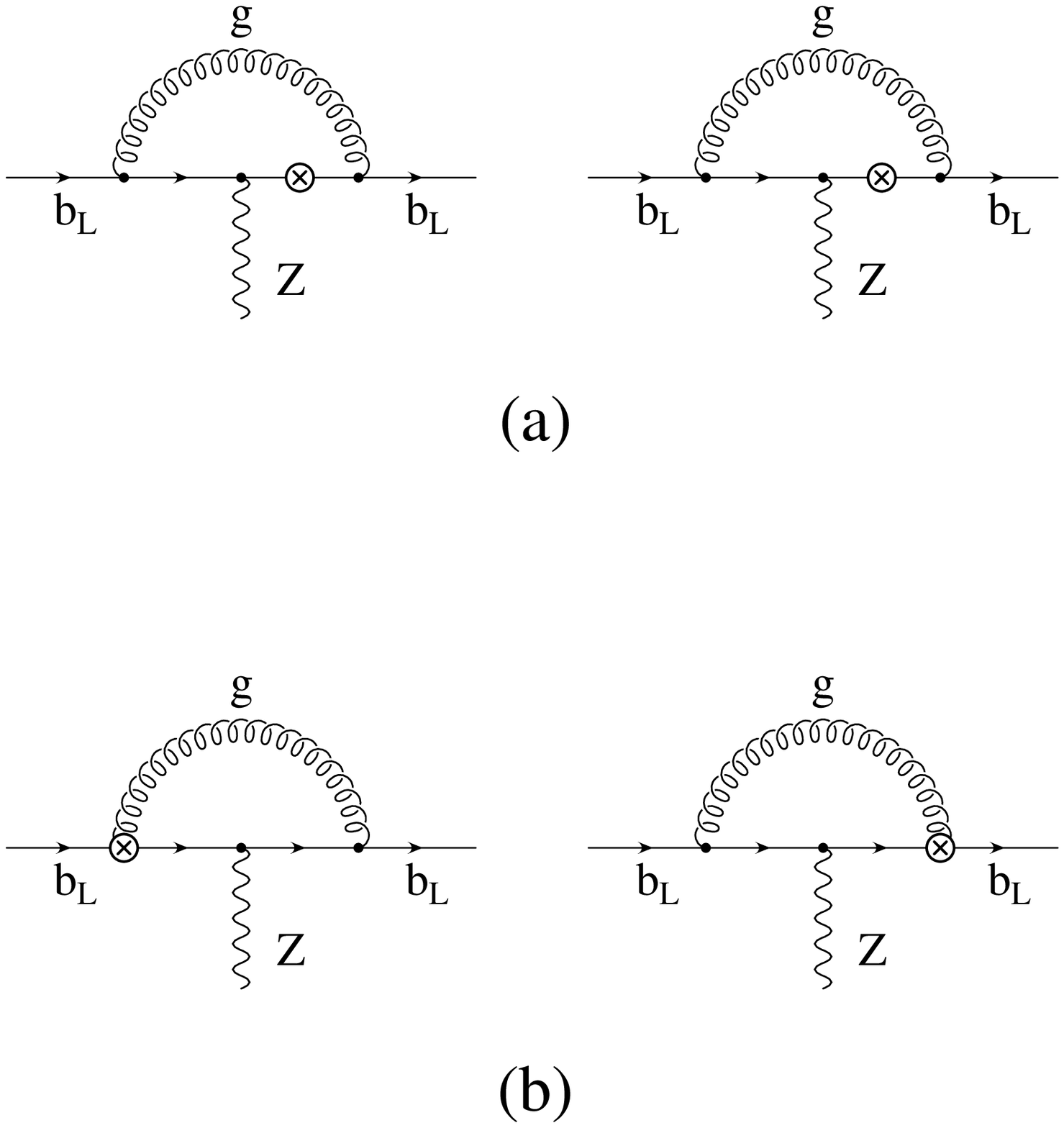}
{QCD running: Insertion of $b$-$b$ and $b$-$b$-$g$ operators.}{zbb7}

Now we would like to integrate eq. (\ref{arca3}). In first
approximation, one may take $\a(t)$ and $c_1(t)$ as constants
independent of $t$, i.e. $\a(\m)\simeq \a(m_t)\simeq \a(M)\equiv \a$
and $c_1(\m)\simeq c_1(m_t)\simeq c_1(M)\equiv c_1 = -7/18$. The
integration over $t$ between $\log m_t^2$ and $\log M^2$ gives

\beq
\label{arca4}
c_L^b(M)\simeq c_L^b(m_t) + \left(\mathrm{something} \right)
\ \log \frac{M^2}{m_t^2} +
\frac{g}{c} \frac{g^2}{\4pi^2}\ \gamma_1 \ c_1\ \frac{\a}{\pi} \log
\frac{M^2}{m_t^2} \ \ .
\eeq
It is in principle possible to improve on this approximation by
considering the $\m$-dependence of $\a(\m)$ and $c_1(\m)$ in eq.
(\ref{arca3}). The
$\m$-dependence of $\a(\m)$ is given by the usual one-loop $\beta$
function. However the $\m$-dependence of $c_1(\m)$ is more complicated
to obtain because it requires performing a complete operator mixing
analysis of the penguin operator along the lines of, for instance,
the work carried
out in the studies of $b\rightarrow s\gamma$ or $K^0-\bar K^0$ mixing
\cite{Martinelli-altres} whence most of the results
could be taken over to our case. However, the fact that
$\gamma_1 c_1(m_t)=\frac{7}{162 c^2}\ (1-\frac{2}{3}\ s^2)\approx 0.05$
turns out to be so small renders this improvement moot and we
shall content ourselves with eq. (\ref{arca4}) as it is. As we shall see
later on, there are other sources of QCD corrections that are numerically
more important.

One obtains (see ref. \cite{Peris-Santamaria} for details):

\beq
\label{arca5}
\frac{c_L^b(M)}{c_L(M)} \approx \frac{c_L^b(m_t)}{c_L(\mt)} \left[1 +
\frac{g^2}{\4pi^2}
\left(\frac{4}{3}+\frac{1}{12c^2}\right)\log\frac{M^2}{m_t^2}+
\frac{g^2}{\4pi^2} \ \gamma_1 \ c_1 \frac{\a}{\pi}\log\frac{M^2}{m_t^2}
\right] \ \ .
\eeq
This fixes the coefficient ${\cal C}$ in eq. (\ref{seven}) to be (remember
eq. (\ref{alta}))

\beq
\label{arcaarca}
{\cal C}= 2\ \gamma_1 \ c_1\ \left(\frac{8}{3}+\frac{1}{6 c^2}\right)^{-1}
\approx 0.03\ \ .
\eeq

The boundary condition at $m_t$ can be borrowed from the literature
\cite{match-mt-2loops}. Translated into our
context it amounts to
\beq
\label{arca6}
\frac{c_L^b(m_t)}{c_L(m_t)} = 1 - 2\ \frac{m_t^2(\mt)}{(4\pi v)^2}\left[1 -
\frac{\a(\mt)}{\pi}\ \left( \frac{\pi^2-8}{3}\right)\right]\ \ .
\eeq

Again, what the EFT tells us is that the $\m$ scale of $\a(\m)$
in this equation  has to be $m_t$ since it originates at the matching
condition when the top is integrated out.\footnote{Another advantage is
that matching conditions are free from infrared divergences\cite{efts},
which is a nice simplification. For
some more discussion on infrared divergences, see below.} Therefore we
get to eq. (\ref{seven}) with $\a(\m=m_t)$ in the $m_t^2$-dependent
term.

\mafigura{13cm}{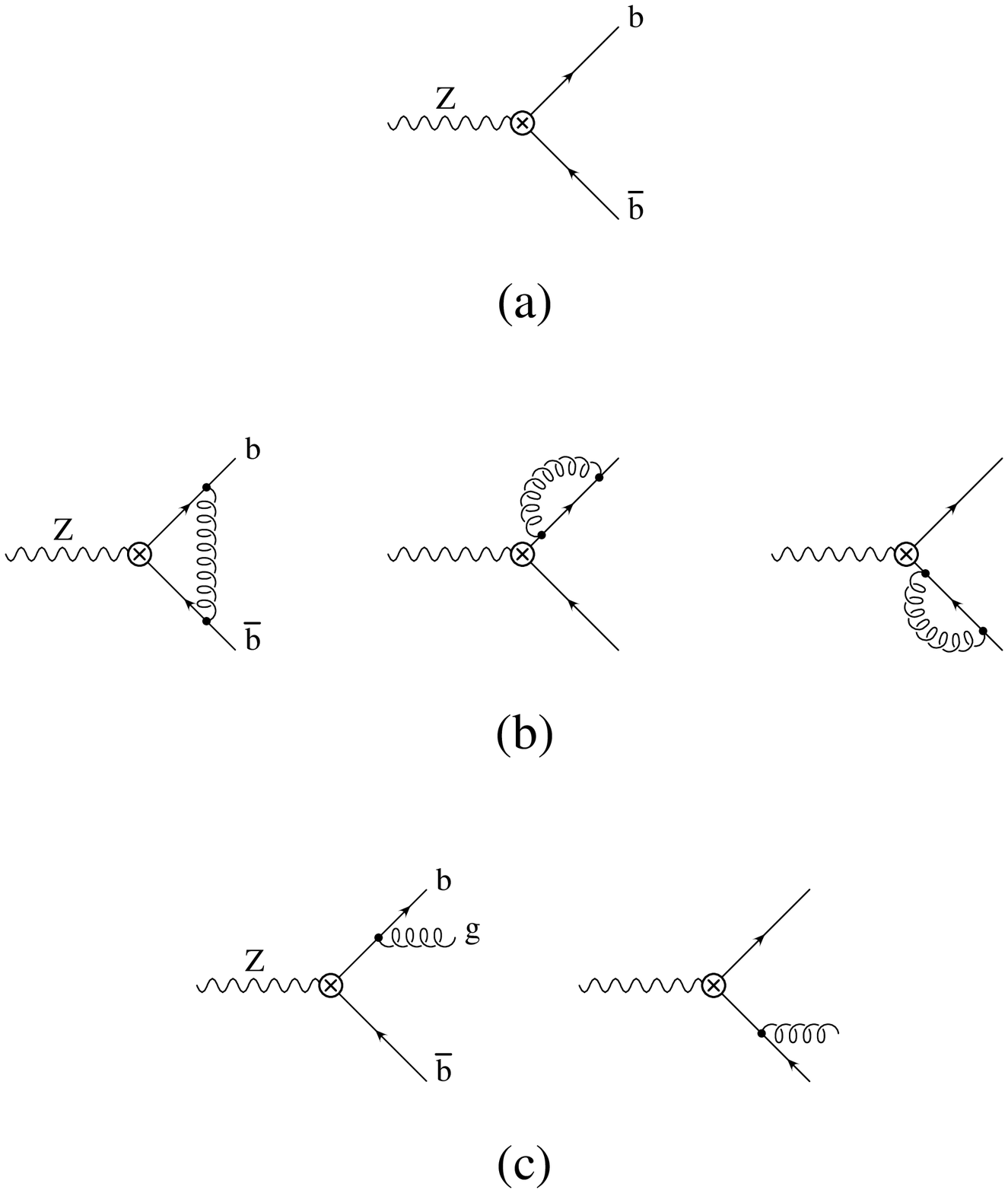}{Diagrams contributing to the matrix element
of $\zbb$ in the effective theory.}{zbb9}

However this is not yet all. Up to now all the physics has been
described with RGEs (i.e. running) and their initial
conditions (i.e. matching)
which is only ultraviolet physics, and
no reference to infrared physics has
been made. For instance, where are the infrared divergences that appear
when a gluon is radiated off a bottom-quark leg? As we shall now see,
this physics is in
the matrix element for $Z\rightarrow b\bar b$. After
all, we have only obtained the effective Lagrangian
(\ref{one2}) at the scale $\mu=M$; we still have to
compute the physical matrix element with it, and here is where all the
infrared physics takes place.

Indeed, when computing the matrix element for $Z\rightarrow b \bar b$
with the effective Lagrangian (\ref{one2}) expressed in
terms of $c_{L,V}^b(\m)$ at $\m=M$, one has the contribution of the
diagrams of figs.~\ref{zbb9}a, \ref{zbb9}b,
where the $\otimes$ stands for the
effective vertices proportional to $c_{L,V}^b(M)$. These diagrams give
rise to infrared divergences. These divergences disappear in the
standard way once bremsstrahlung diagrams like those of fig.~\ref{zbb9}c
are (incoherently) added \cite{IR,bile}.

As is well known \cite{RQCD}, the net result of all this
(a similar calculation can be performed for the QED corrections)
is the appearance of the
factors $R_{QCD}$ and $R_{QED}$ of eqs. (\ref{one}) and (\ref{six}),
where $b$-quark mass effects can also be included
\cite{djouadi-chetyrkin,bile} if needed.

The EFT
technology adds to this the choice of scale for $\m$, namely $\m=M$, in
these factors: \footnote{\ This has been previously suggested by D. Bardin
(private communication).}
\beq
\label{arca7}
R_{QCD}\simeq 1 + \frac{\a(M)}{\pi}\quad , \quad R_{QED} \simeq 1 +
\frac{\alpha(M)}{12\pi}\ \ ,
\eeq
and naturally leads to the factorized expression (\ref{one})--(\ref{seven})
with the value of ${\cal C}$ given by eq. (\ref{arcaarca}). As previously
stated, our result agrees with that of
ref.~\cite{Kwiatkowski-Steinhauser}. Since the
``intrinsic'' $\a$ contribution of $\Delta \rho^{vertex}$ is, due to the
smallness of the coefficient ${\cal C}$, much less important than that
of $R_{QCD}$ one sees that the QCD corrections to the non-universal
$\log m_t$ piece of the $Zb \bar b$ vertex are, to a very good
approximation, of the form one-loop QCD ($m_t<<M_Z$) times one-loop
electroweak ($m_t>>M_Z$)\cite{Peris-unpublished}.

\section{Conclusions}

We hope we have been able to show that the effective field theory construction
can be very useful for multiloop calculations in the Standard Model when
only a few terms in a large mass expansion are needed. Because the
powerfulness of the RGEs is naturally implemented in the EFT, we achieve
full ``logarithmic" control over the relevant scales $\mu$ of the problem
at hand. For instance
the ``large" logarithms are obtained through the beta functions of certain
effective couplings. This only requires the calculation of simple poles in
$1/\epsilon$ which is a major simplification.
We have also shown that the EFT framework answers
quite naturally the question of the renormalization
points to be used for the coupling constants
in the different terms.

In addition it is important to remark that in the
EFT language all the physics
above $M$ is absorbed (in particular, all $\mt$ effects) in the
coefficients of the effective operators so that infrared physics is
relegated to the calculation of the physical process one is
interested in. With our effective Lagrangian
one could in principle compute any physical quantity, and not only
the $Z$ width, like for example jet production (i.e. where cuts are needed)
, forward--backward asymmetries, etc.
This is to be compared with more standard methods in which, in order to
avoid problems with infrared divergences, one computes the imaginary part of
the $Z$ self-energy to obtain the $Z$ width. In this case it is not at all
clear how one can tailor to one's needs the entire phase space.

The EFT calculation clearly separates ultraviolet from infrared physics and
as a consequence it is more flexible. And it is also simpler since, after
all, we never had to compute anything more complicated than a one-loop
diagram.

Of course, our results become more accurate as the top mass becomes larger.
In practice it is unlikely that the top quark be much heavier than, say,
$200$~GeV so due caution is recommended in the phenomenological use of eq.
(\ref{one}), for instance. In the lack of a (very hard !) full ${\cal
O}(g^2\a)$ calculation, this is the best one can offer. Furthermore, we
think it is interesting that at least there exists a limit (i.e. $m_t>>M_Z$)
where the various contributions are under full theoretical control.

\section{Acknowledgements}

I would like to thank A. Santamaria for a most pleasant collaboration and
for multiple interesting discussions. I would also like to thank
B. Kniehl for the nice organization of this workshop and for the
kind invitation. Finally I would like to thank people at the workshop for
interesting conversations, specially K.G.
Chetyrkin, F. Jegerlehner, J.H. K\"uhn, A. Sirlin and M. Steinhauser.

This work was partly supported by CICYT, Spain, under grants AEN93-0474.

\vfil\eject
%

\end{document}